\documentclass[conference]{IEEEtran}
\IEEEoverridecommandlockouts
\usepackage{cite}
\usepackage{amsmath,amssymb,amsfonts,amsthm, empheq}
\usepackage{algorithmic}
\usepackage{graphicx}
\usepackage{textcomp}
\usepackage{xcolor}
\usepackage{hyperref}
\usepackage{booktabs, ragged2e}
\usepackage{multirow}
\usepackage[flushleft]{threeparttable}
\usepackage{makecell}
\usepackage[ruled,vlined,linesnumbered]{algorithm2e}

\theoremstyle{definition}

\newtheorem{lem}{Lemma}

\makeatletter
\newenvironment{proof*}[1][\proofname]{\par
  \pushQED{\qed}%
  \normalfont \partopsep=\z@skip \topsep=\z@skip
  \trivlist
  \item[\hskip\labelsep
        \itshape
    #1\@addpunct{.}]\ignorespaces
}{%
  \popQED\endtrivlist\@endpefalse
}
\makeatother

\makeatletter
\def\thm@space@setup{\thm@preskip=1pt
\thm@postskip=1pt}
\makeatother

\usepackage[letterpaper, left=0.68in, right=0.68in, bottom=1in, top=0.72in]{geometry}
\def\BibTeX{{\rm B\kern-.05em{\sc i\kern-.025em b}\kern-.08em
    T\kern-.1667em\lower.7ex\hbox{E}\kern-.125emX}}

\begin{document}

\title{FAST: Fidelity-Adjustable Semantic Transmission over Heterogeneous Wireless Networks% Energy-Efficient Data Transmission
\vspace*{-0.2cm}}
% \author{\IEEEauthorblockN{Peichun Li,
% Guoliang Cheng, 
% Rong Yu and 
% Yuan Wu
% }}
% \vspace*{-0.8cm}
\author{\IEEEauthorblockN{Peichun Li\IEEEauthorrefmark{1}\textsuperscript{,}\IEEEauthorrefmark{2},
Guoliang Cheng\IEEEauthorrefmark{1}, 
Jiawen Kang\IEEEauthorrefmark{1},
Rong Yu\IEEEauthorrefmark{1},
Liping Qian\IEEEauthorrefmark{3},
Yuan Wu\IEEEauthorrefmark{2},
% Miao Pan\IEEEauthorrefmark{3}
and Dusit Niyato\IEEEauthorrefmark{4}
}
\IEEEauthorblockA{\IEEEauthorrefmark{1}School of Automation, Guangdong University of Technology, Guangzhou, China\\
\IEEEauthorrefmark{2}State Key Laboratory of Internet of Things for Smart City, University of Macau, Macau, China\\
\IEEEauthorrefmark{3}College of Information Engineering, Zhejiang University of Technology, Hangzhou, China\\
% \IEEEauthorrefmark{3}Department of Electrical and Computer Engineering, University of Houston, Houston, USA\\
\IEEEauthorrefmark{4}School of Computer Science and Engineering, Nanyang Technological University, Singapore\\
Email: peichun@mail2.gdut.edu.cn, guoliang\_cheng@126.com, \{kavinkang, yurong\}@gdut.edu.cn, \\
lpqian@zjut.edu.cn, yuanwu@um.edu.mo, dniyato@ntu.edu.sg}
\vspace*{-0.8cm}
}

% mpan2@uh.edu, dniyato@ntu.edu.sg
\maketitle

\begin{abstract}
In this work, we investigate the challenging problem of on-demand semantic communication over heterogeneous wireless networks. We propose a fidelity-adjustable semantic transmission framework (FAST) that empowers wireless devices to send data efficiently under different application scenarios and resource conditions. To this end, we first design a dynamic sub-model training scheme to learn the flexible semantic model, which enables edge devices to customize the transmission fidelity with different widths of the semantic model. After that, we focus on the FAST optimization problem to minimize the system energy consumption with latency and fidelity constraints. Following that, the optimal transmission strategies including the scaling factor of the semantic model, computing frequency, and transmitting power are derived for the devices. 
Experiment results indicate that, when compared to the baseline transmission schemes, the proposed framework can reduce up to one order of magnitude of the system energy consumption and data size for maintaining reasonable data fidelity.
\end{abstract}


\section{Introduction}
By 2030, 17.1 billion wireless devices equipped with versatile sensors will produce 5 zettabytes of data per month \cite{chowdhury20206g}. 
The explosive growth of edge-generated data rises challenges on how to achieve efficient information exchange among massive devices.
Semantic communication is an emerging data transmission paradigm that aims to extract and deliver the explicative meaning of the data \cite{bao2011towards}. 
The semantic-aware communication systems can reveal the intrinsic information of the raw data by leveraging the knowledge of prior models \cite{xie2021deep, 9832831}.
By integrating semantic communication into wireless networks, the required data traffic will be significantly reduced, leading to a green and reliable communication pattern \cite{shi2021semantic, yang2022semantic}. 

By leveraging the capacity of neural networks, learning-based semantic communication systems can extract compact and accurate information from the image and speech \cite{weng2021semantic,luo2022autoencoder, huang2021deep}. To improve the freshness of status updates, the age of semantics is incorporated into the semantic communication systems  \cite{chen2022age}. Recently, system-level methods focus on improving the efficiency of semantic communication, such as the spectrum-efficient method that assigns the optimal channel for the wireless devices \cite{9763856}, and adaptive resource scheduling that maximizes the successful probability of transmission \cite{liu2022adaptable}. 
However, these methods employ fixed neural networks to accomplish the extraction of semantic information during the running time, which hinders the flexibility of semantic-aware transmission over heterogeneous networks.

\begin{figure}\centering
  \includegraphics[width=0.48\textwidth]{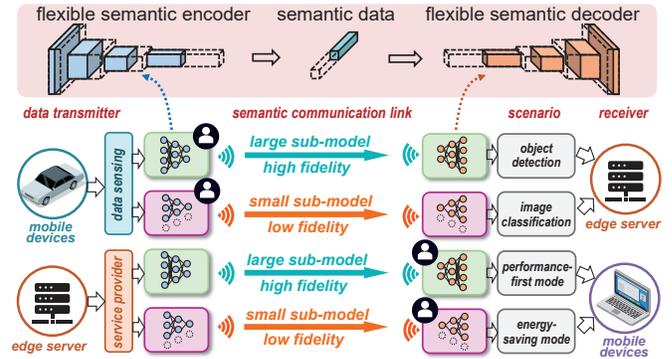}
  \vspace{-8pt}
  \caption{FAST over heterogeneous wireless networks. 
}\label{fig:fast-sys}
\vspace{-14pt}
\end{figure}

Using a fixed learning-based semantic model is a stringent limit for the communication system over heterogeneous wireless networks. 
This setting deteriorates the ability of the communication system in handling different application scenarios. As illustrated in Figure \ref{fig:fast-sys}, a typical semantic model may be designed to concurrently support multiple vision-related tasks under different resource conditions. Compared with image classification that only identifies the category of the image, object detection needs to additionally analyze the location of the object of interest \cite{szegedy2013deep}. Thus, given limited computation and communication resources, high-fidelity semantic data with fine-grained information should be reserved for object detection \cite{song2020fine}, while the low-fidelity one is sufficient for image classification. Also, the quality of semantic communication should be adapted to the energy status of the battery-powered devices. Employing high-fidelity mode for performance-first setting and switching to low-fidelity mode for the purpose of energy saving is an effective way to maintain the transmission quality while prolonging the battery lifetime \cite{yu2021toward, wu2022non}.

In this paper, we propose FAST, a fidelity-adjustable semantic transmission framework, to improve the flexibility of learning-based semantic communication systems over heterogeneous wireless networks. We focus on the image transmission task with autoencoder as the semantic model \cite{tong2021federated}. Our goal is to train a flexible semantic model that enables wireless devices to select different sizes of sub-models at the running time. Then, we propose a fidelity-aware resource management approach, where the optimal transmission strategy is designed to meet the quality and efficiency constraints.
Specifically, a full-size model with powerful capacity is preferred in the high-fidelity scenario, and a small sub-model is adopted in the low-fidelity scenario to save the system cost \cite{li2023anycostfl}.

However, determining the optimal transmission strategies for FAST with personalized constraints is a non-trivial task, as how to train the flexible semantic models and how the fidelity of semantic data is affected by model size is unknown.
To address these issues, we first design a dynamic sub-model training scheme to concurrently support flexible encoding and decoding with different model widths. Meanwhile, the relationship between model size and the expected fidelity is empirically quantified. Following that, we study the FAST optimization problem to improve energy efficiency with given latency and fidelity budgets. Based on the theoretical analysis, the problem is transformed into a convex problem. Finally, we develop a hierarchical bisection algorithm to solve the problem, where the size of the semantic sub-model, CPU computing frequency, and transmitting power are determined according to the fidelity constraint and resource status.

Our main contributions are summarized as follows.
\begin{itemize}
	\item We propose a novel semantic communication framework, named FAST that enables wireless devices to perform fidelity-adjustable data transmission.
	\item We investigate the fidelity-aware resource management problem for FAST, and the optimal transmission strategy is devised to minimize the system energy cost.
	\item Extensive experiments demonstrate the efficiency and effectiveness of FAST, which outperforms the existing baselines in terms of resource utilization and data fidelity.
\end{itemize}

The remainder of this paper is organized as follows. Section II details the main components of FAST to fulfill flexible semantic communication. 
The problem formulation, theoretical analysis, and the corresponding solution are provided in Section III. 
The experiment simulations are presented in Section IV, and we finally conclude the paper in Section V.

\section{System Model}

\subsection{Outline of FAST}
\begin{figure}\centering
  \includegraphics[width=0.48\textwidth]{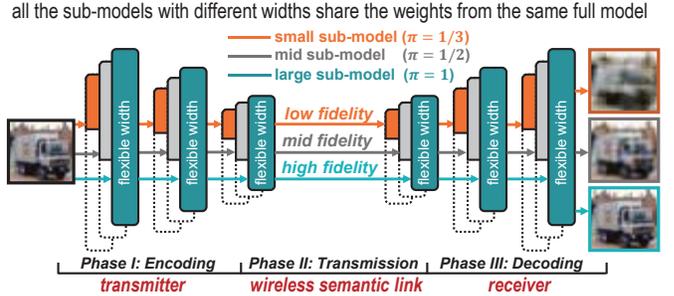}
  \vspace{-8pt}
  \caption{FAST with flexible semantic model. 
}\label{fig:fast-model}
  \vspace{-12pt}
\end{figure}

We consider the scenario of wireless semantic communication between two physical entities, i.e., the transmitter and the receiver. As shown in Figure~\ref{fig:fast-model}, unlike traditional methods that utilize the fixed model during the running time, FAST employs a flexible semantic model to accomplish the fidelity-adjustable transmission. 
Specifically, the semantic model comprises of two parts, including the encoder and decoder. For the full-size model, we use $\boldsymbol{\theta} $ and $\boldsymbol{\vartheta} $ to parameterize the weights of the encoder and decoder, respectively.
Here, we introduce a scaling factor $\pi\in(0,1]$ for the width of each layer in the flexible model. Given a scaling factor $\pi$, we can derive a pair of small encoder and decoder from the full-size model, denoted as $\boldsymbol{\theta}_{\pi}$ and $\boldsymbol{\vartheta}_{\pi}$, respectively. 
The process of the FAST is divided into the following three phases.
\subsubsection{Phase I for encoding} With the pre-determined scaling factor $\pi$, the source device switches from the full-size encoder to a small one parameterized by $\boldsymbol{\theta}_{\pi}$. Let $\boldsymbol{x}$ denote the raw data, and let ${\boldsymbol{h}}_{\pi}$ represent the corresponding semantic data. The function of semantic encoding can be expressed as
\begin{equation}\label{eqn:encode}
{\boldsymbol{h}}_{\pi} = \texttt{enc}({\boldsymbol{x}}; \boldsymbol{\theta}_{\pi}).
\end{equation}

\subsubsection{Phase II for transmission} After obtaining the semantic data ${\boldsymbol{h}}_{\pi}$, the source device transmits it to the destination. Here, we consider that the semantic data is converted into binary symbols. Thus, the transmission for the semantic information still follows the Shannon capacity \cite{liu2022adaptable}.
\subsubsection{Phase III for decoding} After receiving the semantic data, the destination device decodes it to reconstruct the data $\hat{{\boldsymbol{x}}}$. Specifically, the encoder and decoder have the symmetry structures, i.e., the decoder shares the same scaling factor $\pi$ as the encoder does. The process of semantic decoding can be represented as
\begin{equation}\label{eqn:decode}
\hat{{\boldsymbol{x}}} = \texttt{dec}({\boldsymbol{h}}_{\pi}; \boldsymbol{\vartheta}_{\pi}).
\end{equation}

\vspace{-5pt}
\subsection{Flexible Semantic Model}
We aim to train a flexible semantic model that supports nearly continuous scaling factor $\pi \in (0,1]$. We first focus on deriving a pair of small-size encoder and decoder $\{ \boldsymbol{\theta}_{\pi},  \boldsymbol{\vartheta}_{\pi}\}$ from $\{ \boldsymbol{\theta},  \boldsymbol{\vartheta}\}$. Then, we propose a dynamic sub-model training scheme to train the flexible semantic model efficiently. 
\subsubsection{Sub-model derivation}
Given a scaling factor $\pi$, we aim to derive the small-size sub-model $\{\boldsymbol{\theta}_{\pi}, \boldsymbol{\vartheta}_{\pi}\}$ from the full-size one $\{ \boldsymbol{\theta},  \boldsymbol{\vartheta}\}$.
Specifically, the sub-model derivation is performed in a layer-by-layer manner. For a convolution layer with a number of filters as $C$ (i.e., the width of the layer), we select the weights of the first $\lfloor \pi C \rfloor$ filters to construct the layer for a small-size sub-model. Given a scaling factor $\pi$, the derivations for $\boldsymbol{\theta}_{\pi}$ and $\boldsymbol{\vartheta}_{\pi}$ are respectively expressed by
\begin{equation}\label{eqn:derivation}
\boldsymbol{\theta}_{\pi} = \texttt{sel}(\boldsymbol{\theta},{\pi}) \quad\textrm{and \quad} \boldsymbol{\vartheta}_{\pi} = \texttt{sel}(\boldsymbol{\vartheta},{\pi}),
\end{equation}
where $\texttt{sel}(\cdot,\cdot)$ denotes the function to select the weight from the full-size model to the small one.

\subsubsection{Learning objective} Our goal is to minimize the data reconstruction error for any sub-models derived from the full-size model. Let $\ell(\boldsymbol{x}, \hat{\boldsymbol{x}})$ be the pre-determined loss function. The learning objective can be expressed as
\begin{equation}\label{eqn:learn-obj}
\mathop {\min }\limits_{\boldsymbol{\theta},\boldsymbol{\vartheta}} \int_{\pi _{\min}}^{1} \sum\limits_{\boldsymbol{x} \in \boldsymbol{X}_{\textrm{test}}}{\ell \big({\boldsymbol{x}}, F({\boldsymbol{x}};\boldsymbol{\theta},\boldsymbol{\vartheta},\pi) \big)} d\pi,
\end{equation}
where the function $F({\boldsymbol{x}};\boldsymbol{\theta},\boldsymbol{\vartheta},\pi) = \texttt{dec}\big(\texttt{enc}({\boldsymbol{x}}; \boldsymbol{\theta}_{\pi}); \boldsymbol{\vartheta}_{\pi}\big)$ computes the reconstructed data with given $\boldsymbol{x},\boldsymbol{\theta},\boldsymbol{\vartheta}$ and $\pi$.

\subsubsection{Dynamic sub-model training} 
Note that the process of optimizing Eqn.~(\ref{eqn:learn-obj}) requires enumerating all sub-models, which incurs a prohibitive cost for the model training. Inspired by the study in \cite{lin2021anycost}, an efficient training method via sub-model sampling is proposed to reduce the computation cost. As presented in Algorithm \ref{alg:train}, we propose to dynamically sample a sub-model at each iteration (i.e., Steps 3). Specifically, the total training loss is computed as the sum of the losses of the random sub-model and the full-size model (i.e., Step 5). In this way, we maintain the performance of different sub-model while reducing the training overhead.

\begin{algorithm}
\SetAlgoLined
\SetKwInput{KwInput}{Input}
\SetKwInput{KwOutput}{Output}
\SetKwRepeat{KwRepeat}{repeat}{until}
\KwInput{Training dataset $\boldsymbol{X}_{\textrm{train}}$, and $\{ \boldsymbol{\theta},\boldsymbol{\vartheta}\}$.}
\For{\normalfont{each epoch} $i=1,2,\ldots,I$}{
    \For{\normalfont{each batch of training data} $\boldsymbol{x}_{\textrm{batch}}\in\boldsymbol{X}_{\textrm{train}}$}{
        Randomly sample a scaling factor $\pi$.
        Perform forward propagation with $\pi$: $\hat{\boldsymbol{x}}_{\textrm{sub}} = F(\boldsymbol{x}_{\textrm{batch}};\boldsymbol{\theta},\boldsymbol{\vartheta},\pi)$\;
        Perform forward propagation with full-size model: $\hat{\boldsymbol{x}}_{\textrm{full}} = F(\boldsymbol{x}_{\textrm{batch}};\boldsymbol{\theta},\boldsymbol{\vartheta}, 1)$\;
        Compute the total reconstruction loss: $Loss=\ell(\boldsymbol{x}_{\textrm{batch}}, \hat{\boldsymbol{x}}_{\textrm{sub}}) + \ell(\boldsymbol{x}_{\textrm{batch}}, \hat{\boldsymbol{x}}_{\textrm{full}})$\;
        Apply backward propagation to update $\{ \boldsymbol{\theta},\boldsymbol{\vartheta}\}$\;
    }
}
\caption{Dynamic sub-model training}
\label{alg:train}
\vspace{-2pt}
\end{algorithm}

\vspace{-8pt}
\subsection{Characterizing the Semantic Fidelity}
We next investigate how the scaling factor $\pi$ affects the performance of the corresponding sub-model. 
We first present the definition of semantic fidelity of the given semantic model, and then reveal the relationship between the scaling factor $\pi$ and the semantic fidelity of the sub-model.

\subsubsection{Definition of semantic fidelity} 
We define the semantic fidelity of the semantic model as the capability of reconstructing data  over the testing dataset. Formally, given a scaling factor of $\pi$, the semantic fidelity $\phi _{\pi}$ of the corresponding sub-model is calculated by
\begin{equation}\label{eqn:sem-fidel}
\phi _{\pi} = 1 - \frac{1}{M|\boldsymbol{X}_{\textrm{test}}|}\sum\limits_{\boldsymbol{x} \in \boldsymbol{X}_{\textrm{test}}}{{\|{\boldsymbol{x}} - F({\boldsymbol{x}};\boldsymbol{\theta},\boldsymbol{\vartheta},\pi)\|}},
\end{equation}
where $M$ denotes the number of pixels in the image, $|\boldsymbol{X}_{\textrm{test}}|$ measures the number of samples of the testing dataset, and $\|\cdot\|$ calculates the L1 norm for the given vector.
\subsubsection{The impact of the scaling factor on semantic fidelity} 
\begin{figure}\centering
  \includegraphics[width=0.45\textwidth]{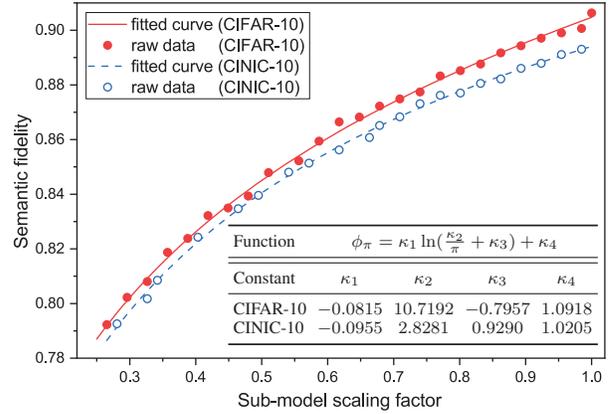}
  \vspace{-10pt}
  \caption{Semantic fidelity $\phi _{\pi}$ of sub-model with respect to scaling factor $\pi$. 
}\label{fig:fitted}
  \vspace{-12pt}
\end{figure}

Intuitively, larger sub-models with strong representation capabilities can extract more latent information from the raw data, resulting in higher semantic fidelity. Being consistent with the existing studies in \cite{li2021fedgreen}, we employ the parameter fitting method to empirically investigate the relationship between semantic fidelity $\phi _{\pi}$ and the scaling factor $\pi$. We adopt the CIFAR-10 and CINIC-10 datasets, and the experiment settings are provided in Section IV. Then, we sample a subset of sub-models from the flexible semantic model trained by Algorithm \ref{alg:train}, and evaluate their corresponding performance by Eqn.~(\ref{eqn:sem-fidel}). The relationship between $\phi _{\pi}$ and $\pi$ is formulated as  
\begin{equation}\label{eqn:fit}
\phi _{\pi} = \kappa _1 \ln(\frac{\kappa _2}{\pi}+\kappa _3)+\kappa _4,
\end{equation}
where $\{\kappa _1, \kappa _2, \kappa _3, \kappa _4\}$ are constant hyper-parameters that can be experimentally fitted. The experiment results are provided in Figure~\ref{fig:fitted}. As $\pi$ increases, the fidelity of semantic data increases to carry more detailed information.

\section{Problem and Solution}
% In this section, we study the energy-efficient transmission problem for our framework. We aim to minimize the system energy consumption of FAST with latency and fidelity constraints. Following that, the optimization problem is transformed into a tractable form, and a hierarchical bisection algorithm is proposed to solve the problem.

\subsection{FAST over Wireless Networks}
For semantic communication with the full-size model and single image sample, we use $W_{e}$ and $W_{d}$ to denote the computation workloads for encoding and decoding, respectively, and the size of the semantic information is $S$. For FAST with a scaling factor of $\pi$, the encoding workloads, decoding workloads, and data size to be transmitted are reduced as $\pi ^2 W_{e}$, $\pi ^2 W_{d}$ and $\pi S$, respectively.

\subsubsection{Computation model}
Let $f_{e}$ and $f_{d}$ denote the computing frequency for the encoding and decoding, respectively.
For the semantic-based transmission with $K$ samples, given the sub-model scaling factor $\pi$, the overall time taken for the model inference can be measured by
\begin{equation}\label{eqn:infer-time}
T_{\text{cmp}} = K\pi ^2\bigg(\frac{{W_{e}}}{f_{e}} + \frac{{ W_{d}}}{f_{d}}\bigg).
\end{equation}
Meanwhile, the overall energy consumption is estimated by
\begin{equation}\label{eqn:update-energy}
E_{\text{cmp}} = K\pi ^2(\epsilon _{e} f_{e}^2 {W_{e}} + \epsilon _{d} f_{d}^2 {W_{d}}),
\end{equation}
where $\epsilon _{e}$ and $\epsilon _{d}$ are the hardware energy coefficients of the source and destination devices, respectively.

\subsubsection{Communication model} Let $B$ denote the available bandwidth, $P$ the transmitting power of the source device and $N_0$ be the power spectral density of the Gaussian noise. For the transmission of semantic data from the source to the destination, the achievable transmitting rate is estimated by
\begin{equation}\label{eqn:trans-rate}
r = {B}{\log _2}\Big(1 + \frac{{|h|^2d^{-\eta}P}}{{{N_0}{B}}}\Big),
\end{equation}
where $d$ represents the distance between the transmitter and receiver, $\eta$ is the pathloss exponent, and $h$ denotes the Rayleigh channel coefficient.
With given scaling factor of $\pi$, the required time $T_{\text{com}}$ and energy consumption $E_{\text{com}}$ for the transmission of semantic data can be respectively calculated by
\begin{equation}\label{eqn:trans-time}
T_{\text{com}} = \frac{K\pi S}{r},~\text{and }
E_{\text{com}} = PT_{\text{com}}.
\end{equation}

\subsubsection{Problem formulation}
% To optimize FAST, we study an energy-efficient optimization problem with personalized constraints. 
Given a pair of source-destination devices with different local resources, our goal is to optimize the transmission strategy for these two devices to minimize the total energy cost with latency and fidelity constraints. To this end, we formulate the following optimization problem.
\begin{subequations}\label{eqn:pbm-1}
\begin{align}
    ({\text{P1}}) && \min&  \; E_{\text{tot}}   &\tag{\ref{eqn:pbm-1}}\\
    {\text{subject to:}} && T_{\text{tot}}  \le &\;{T^{\max }},&\label{eqn:p1-ctr-1}\\
    {} && \phi _{\pi} \ge&\; \phi^{{\min}},&\label{eqn:p1-ctr-2}\\
    {}&& {\pi^{\min}} \le&\; \pi  \le 1, & \label{eqn:p1-ctr-3}\\
    {}&& 0 \le f_{e} \le f_{e} ^{\max}&,\; 0\le f_{d} \le f_{d} ^{\max},& \label{eqn:p1-ctr-4}\\
    {}&& 0 \le P &{}\le P ^{\max},& \label{eqn:p1-ctr-5}
    \\
	{{\textrm{variables:}}}&& \pi,\; f_e&, \;f_d, \;P,&\nonumber
\end{align}
\end{subequations}
where $E_{\text{tot}} = E_{\text{cmp}} + E_{\text{com}}$ and $T_{\text{tot}} = T_{\text{cmp}} + T_{\text{com}}$ are the total energy cost and the total system latency, respectively.
% Specifically, Constraint (\ref{eqn:p1-ctr-1}) represents the total latency cannot exceed $T^{\max}$. Constraint (\ref{eqn:p1-ctr-2}) guarantees the minimum semantic fidelity requirement of $\phi ^{\min}$. Constraints (\ref{eqn:p1-ctr-3})-(\ref{eqn:p1-ctr-5}) restrict the lower and upper limits for the decision variables.

\subsection{Problem Simplification}
In this subsection, we transform Problem (P1) into a tractable yet equivalent form via constraint simplification and variable substitution. We first derive the following lemma.

\begin{lem}
The equality always holds for Constraints (\ref{eqn:p1-ctr-1}) and (\ref{eqn:p1-ctr-2}) under the optimal solution $\{\pi^\ast, f_e^\ast, f_d^\ast, P^\ast\}$, namely, we always have $T_{\text{tot}}^\ast = {T^{\max }}$ and $\phi _{\pi}^\ast = \phi^{{\min}}$.
\label{lem:lem-1}
\end{lem}
\begin{proof*}
We prove the lemma by showing contradictions. Suppose that there exists an optimal solution such that $T_{\text{tot}}^\ast < T^{\max}$. 
Then, we construct a new solution by replacing $f_e^\ast$ with $f_e^\prime$ in the optimal solution such that $f_e^\prime < f_e^\ast$ and  $T_{\text{tot}}^\prime  = T^{\max}$. Let $E_{\text{tot}}^\prime$ denote the corresponding system energy cost of the new solution. 
Since the system energy cost decreases with the decrease of $f_e$, then we obtain $E_{\text{tot}}^\prime < E_{\text{tot}}^\ast$. Similarly, the contradiction also applies for $\phi _{\pi} > \phi ^{\min}$, and thus we complete the proof.
\end{proof*}

Based on Lemma \ref{lem:lem-1} and Eqn.~(\ref{eqn:fit}), we can derive the optimal width scaling factor $\pi ^\ast$ as
\begin{equation}\label{eqn:opt-pi}
\pi ^{\ast} = \frac{\kappa _2}{\exp{\big(\frac{\phi ^{\min} - \kappa_4}{\kappa _1}\big)} - \kappa_3}.
\end{equation}
Moreover, we introduce three intermediate variables $\alpha>0, \beta>0$, and $\gamma>0$ such that $\alpha+ \beta+\gamma=1$, and they denote the time splitting factors of the latency for encoding, semantic transmission, and decoding, respectively. Thus, we obtain the following equations.
\begin{align}
\small
\begin{split}\label{eqn:time-split}
\alpha T^{\max} = K(\pi^\ast) ^2\frac{{W_{e}}}{f_{e}}&, \quad \beta T^{\max} = \frac{K\pi^\ast S}{r},\\\gamma T^{\max} =&\, K(\pi^\ast) ^2\frac{{W_{d}}}{f_{d}}.
\end{split}
\end{align}
By combining Eqns.~(\ref{eqn:infer-time})-(\ref{eqn:trans-time}) and (\ref{eqn:time-split}), the total system energy cost $E_{\text{tot}}$ can be re-expressed with respect to $\{\alpha, \beta, \gamma\}$ as
\begin{equation}\label{eqn:energy}
E_{\text{tot}} = \frac{\tau _1}{\alpha ^2} + \tau _2 \beta \big( 2^{\frac{\tau _3}{\beta}}-1 \big) + \frac{\tau _4}{\gamma ^2},
\end{equation}
where the constants $\{\tau _1, \tau _2, \tau _3, \tau _4\}$ can be calculated by 
\begin{align}
\small
\begin{split}\label{eqn:tau}
\tau _1 = {\epsilon _e}\frac{{K^3W_e^3{(\pi^\ast) ^6}}}{{{{\left( {{T^{\max }}} \right)}^2}}}>0,\;\; \tau _2 = \frac{{B{N_0}{T^{\max }}}}{|h|^2d^{-\eta}}>0,\\ \tau _3 = \frac{{K\pi^\ast S}}{{B{T^{\max }}}}>0,\;\;\tau _4 = {\epsilon _d}\frac{{K^3W_d^3{(\pi^\ast) ^6}}}{{{{\left( {{T^{\max }}} \right)}^2}}}>0.
\end{split}
\end{align}

Therefore, given the optimal scaling factor $\pi^\ast$, Problem (P1) can be transformed into the following problem.
\begin{subequations}\label{eqn:pbm-2}
\begin{align}
    ({\text{P2}}) && \min\;   \frac{\tau _1}{\alpha ^2} + \tau _2 \beta &\big( 2^{\frac{\tau _3}{\beta}}-1 \big) + \frac{\tau _4}{\gamma ^2}   &\tag{\ref{eqn:pbm-2}}\\
    {\text{subject to:}} && \alpha+ \beta\,+ & \,\gamma=1,&\label{eqn:p2-ctr-1}\\
    {} &&  \alpha ^{\min}\le \alpha,\; \beta ^{\min}&\le \beta,\; \gamma ^{\min}\le \gamma,&\label{eqn:p2-ctr-2}\\
    {{\textrm{variables:}}}&& \alpha, \beta,&\; \gamma,&\nonumber
\end{align}
\end{subequations}
where the lower limits of $\{\alpha, \beta, \gamma\}$ can be acquired by
\begin{align}\small
\begin{split}\label{eqn:lower-limit}
\alpha ^{\min} = \frac{K(\pi^\ast) ^2{W_{e}}}{f_{e}^{\max}T^{\max}},\; \gamma^{\min} \,= \frac{K(\pi^\ast) ^2{W_{d}}}{f_{d}^{\max}T^{\max}}, \\
\beta ^{\min} = \frac{K\pi^\ast S}{T^{\max}{B}{\log _2}\Big(1 + \frac{{|h|^2d^{-\eta}P^{\max}}}{{{N_0}{B}}}\Big)}.
\end{split}
\end{align}
Notably, the optimal solution of Problem (P1) can be obtained directly with the help of $\{\alpha^{\ast}, \beta^{\ast}, \gamma^{\ast}\}$ according to Eqn. (\ref{eqn:time-split}). It can be verified that Problem (P2) is a convex optimization problem, and we discuss the solution in next subsection.

\subsection{Hierarchical Bisection Search}
To solve Problem (P2), we first apply Karush–Kuhn–Tucker (KKT) conditions to derive necessary equations for achieving the optimality. By utilizing $\lambda$ as the Lagrange multiplier for the equality Constraint (\ref{eqn:p2-ctr-1}), and $\{\mu _{\alpha}, \mu _{\beta}, \mu _{\gamma}\}$ as the multipliers for the inequality Constraint (\ref{eqn:p2-ctr-2}), we obtain
\begin{subequations}\small
\begin{empheq}[left=\empheqlbrace]{align}
&{\mu _{\alpha}}={\frac{{ - 2{\tau _1}}}{{{\alpha ^3}}} + \lambda}, \,{\mu _{\gamma}} = {\frac{{ - 2{\tau_4}}}{{{\gamma ^3}}} + \lambda}, \label{eqn:p2-kkt-1}\\
&{\mu _{\beta}}={\big( {{\tau_2} - \frac{{{\tau_2}{\tau_3}\ln 2 }}{\beta }} \big){2^{\frac{{{\tau_3}}}{\beta }}} - {\tau_2} + \lambda},\label{eqn:p2-kkt-2}\\
&{\mu _{\alpha}}( \alpha-{{\alpha ^{\min }} } ) = {\mu _{\beta}}( \beta- {{\beta ^{\min }} } ) = {\mu _{\gamma}}( \gamma -{{\gamma ^{\min }}  } ) = 0,\label{eqn:p2-kkt-3}\\
&0\le {\mu _{\alpha}},0\le{\mu _{\beta}},0\le{\mu _{\gamma}},\label{eqn:p2-kkt-4}\\
&\text{Constraints~(\ref{eqn:p2-ctr-1}) and (\ref{eqn:p2-ctr-2})}. \nonumber
\end{empheq}
\end{subequations}

By substituting $\{\mu_{\alpha}, \mu_{\beta}, \mu_{\gamma}\}$ from Eqns~(\ref{eqn:p2-kkt-1}) and (\ref{eqn:p2-kkt-2}) into Eqn.~(\ref{eqn:p2-kkt-3}), we have
\begin{align} \small
\begin{split}
&\Big({\frac{{ - 2{\tau _1}}}{{{\alpha ^3}}} + \lambda}\Big)(\alpha - \alpha^{\min})=  \Big({\frac{{ - 2{\tau_4}}}{{{\gamma ^3}}} + \lambda}\Big)(\beta - \beta^{\min})=0, \label{eqn:p2-kkt-b}
\end{split}\\
\begin{split}
    \Big(\underbrace{{\big( {{\tau_2} - \frac{{{\tau_2}{\tau_3}\ln 2 }}{\beta }} \big){2^{\frac{{{\tau_3}}}{\beta }}} - {\tau_2} + \lambda}}_{g_{\lambda}(\beta)}\Big)(\gamma - \gamma ^{\min})=0.\label{eqn:p2-kkt-c}
\end{split}
\end{align}
According to Constraint (\ref{eqn:p2-ctr-2}), the discussion on the value of $\alpha ^\ast$ can be divided into two cases, i.e., $\alpha ^\ast > \alpha^{\min}$ and $\alpha ^\ast = \alpha ^{\min}$. Based on Eqn.~(\ref{eqn:p2-kkt-b}), we have $\alpha ^\ast= \sqrt[3]{{\frac{{2{\tau _1}}}{\lambda }}} $ if $\alpha ^\ast > \alpha^{\min}$. Similarly, the optimal values of $\{\beta^\ast, \gamma ^\ast\}$ can be analyzed on the same basis. Therefore, we have
\begin{align}\small\begin{split}\label{eqn:opt-by-lambda}
\alpha ^\ast_{\lambda} = \max\big\{\sqrt[3]{{\frac{{2{\tau _1}}}{\lambda }}}, \alpha^{\min}\big\}&{},\;\beta^\ast_{\lambda} = \max\{\beta_{\lambda},\beta^{\min}\},\\
\gamma ^\ast_{\lambda} = \max\big\{&{}\sqrt[3]{{\frac{{2{\tau _4}}}{\lambda }}}, \gamma^{\min}\big\},
\end{split}
\end{align}
where $\beta _{\lambda}$ is the zero of function $g_{\lambda}(\beta)$ in Eqn.~(\ref{eqn:p2-kkt-c}) such that $g_{\lambda}(\beta_{\lambda})=0$.
Given a specific $\lambda$, we define that $z_{\lambda}=\alpha^\ast_{\lambda} + \beta ^\ast_{\lambda} + \gamma ^\ast_{\lambda}$. According to Constraint (\ref{eqn:p2-ctr-1}), the solution of Problem (P2) can be acquired by searching an optimal Lagrange multiplier $\lambda^\ast$ such that $z_{\lambda^\ast}=1$.

It can be verified that $\alpha^\ast_{\lambda}$, $\beta^\ast_{\lambda}$ and $\gamma^\ast_{\lambda}$ are monotonically non-increasing with respect to $\lambda$. Hence, $\lambda^\ast$ can be efficiently obtained by the bisection search as shown in Algorithm \ref{alg:bin-search}. Specifically, given a $\lambda$, $\alpha ^\ast_{\lambda}$ and $\gamma ^\ast_{\lambda}$ be directly calculated while $\beta ^\ast_{\lambda}$ involves another bisection search (i.e., Steps 8-12). Given the tolerance value $\varepsilon$ and searching range $\{\lambda ^{\min}, \lambda ^{\max}, \beta ^{\min}, \beta ^{\max}\}$ and $J=\max\{\lambda^{\max} - \lambda^{\min}, \beta^{\max} - \beta^{\min}\}$, the computational complexity of Algorithm \ref{alg:bin-search} is estimated by ${\cal O}(log_2^2J)$.
% the number of iterations to obtain $\lambda ^\ast$ is estimated by $\log_2{\frac{\lambda^{\max} - \lambda^{\min}}{\varepsilon}}\log_2{\frac{\beta^{\max} - \beta^{\min}}{\varepsilon}}$. Formally,
\vspace{-5pt}
\begin{algorithm}
\SetAlgoLined
\SetKwInput{KwInput}{Input}
\SetKwInput{KwOutput}{Output}
\SetKwRepeat{KwRepeat}{repeat}{until}
\KwInput{$\lambda ^{\min}, \lambda ^{\max}, \beta ^{\min}, \beta ^{\max}$, and $\varepsilon$.}
\KwOutput{The optimal Lagrange multiplier $\lambda^\ast$.}
\KwRepeat{$|\lambda^{\max} - \lambda^{\min}| \leq \varepsilon$}{
    $\lambda = (\lambda^{\max} + \lambda^{\min})/2$\;
    Compute $\alpha ^\ast_{\lambda}$ and $\gamma ^\ast_{\lambda}$ based on Eqn.~(\ref{eqn:opt-by-lambda})\;
    Search for $\beta ^\ast_{\lambda}$, and compute $z_{\lambda}=\alpha ^\ast_{\lambda}+\beta ^\ast_{\lambda}+\gamma ^\ast_{\lambda}$\;
    \textbf{if} $z _{\lambda} < 1$ \textbf{then} $\lambda^{\max} = \lambda$ \textbf{else} $\lambda^{\min} = \lambda$\;
}
\textbf{return} $\lambda^\ast$\\
\tcc{Function for searching $\beta ^\ast_{\lambda}$.}
\KwRepeat{$|\beta^{\max} - \beta^{\min}| \leq \varepsilon$}{
    $\beta = (\beta^{\max} + \beta^{\min})/2$\;
    Compute $g_{\lambda}(\beta)$ based on Eqn.~(\ref{eqn:p2-kkt-c})\;
    \textbf{if} $g_{\lambda}(\beta) > 0$ \textbf{then} $\beta^{\max} = \beta$ \textbf{else} $\beta^{\min} = \beta$\;
}
\textbf{return} $\beta^\ast_{\lambda}$\\
\caption{Hierarchical bisection search}
\label{alg:bin-search}
\end{algorithm}
\vspace{-5pt}

\section{Simulation Results}
\subsection{Experiment Settings}
We consider the semantic communication for image transmission with CIFAR10 dataset. 
For the semantic model, we use two three-layer convolutional neural networks with kernel size as 4, stride as 2, and padding as 1 for the encoder and decoder. Specifically, the widths of the encoder and decoder are \{12, 24, 32\} and \{24, 12, 3\}, respectively. The semantic data is represented by $32\times4\times4=512$ numbers in 8-bit unsigned integer and $S=4096$ bits. For training hyper-parameters, the batch size, total epochs and $\pi _{\min}$ are set as 16, 30 and 0.25, respectively. The hyper-parameters for communication $\{B, |h|^2, d, \eta, N_0\}$ are set as \{1MHz, 10$^{-3}$W, 200m, 3.76, $-95$dBm/MHz\} by default. The hyper-parameters for computation $\{W_e, W_d, \epsilon_e, \epsilon_d, K\}$ are empirically set as \{0.65 MCycles, 3.25MCycles, $1^{-26}$, $1^{-26}$, 512\} by default.

\begin{table}\small
  \caption{Performance comparison between FAST and baseline methods on image transmission with CIFAR-10 datasets.}
   \vspace{-5pt}
  \label{tab:compare-exp}
  \centering
  \scriptsize
  \setlength{\tabcolsep}{2.5pt}
  \begin{tabular}{lcccccc}
    \toprule
         Method & \makecell{Data size \\(Mbit)} & \makecell{Comp. Cost \\(GFLOPs)} & \makecell{$E_{\text{cmp}}$ (J)} & \makecell{$E_{\text{com}}$ (J)} & \makecell{$E_{\text{tot}}$ (J)}& \makecell{Fidelity}\\
    \midrule
Raw	& 12.58 (1$\times$) 	& 0 	& 0 	& 2.24 & 2.24 &1\\
JPEG	& 2.76 (4.56$\times$) 	& \textemdash 	&  \textemdash	& \textemdash & \textemdash &0.73\\
\midrule
Prune ($\rho$=0.3)	& 2.31(5.5$\times$) 	& 3.31 	& 1.65 	& 0.53 	& 2.18 	& 0.80\\
Quant (3 bits)	& 1.18 (10.7$\times$) 	& 3.31 	& 1.44 	& 0.26 	& 1.70 	& 0.80\\
FAST ($\pi$=0.3)& \textbf{0.92} (13.7$\times$) 	& \textbf{0.97} 	& 0.01 	& 0.10 	& \textbf{0.11} 	& 0.80\\
\midrule
Prune ($\rho$=0.1)	& 2.71(4.6$\times$) 	& 3.31 	& 1.73 	& 0.64 	& 2.37 	& 0.85\\
Quant (4 bits)	& \textbf{1.57} (8.0$\times$) 	& 3.31 	& 1.51 	& 0.35 	& 1.86 	& 0.85\\
FAST ($\pi$=0.5)	& 1.67 (7.5$\times$) 	& \textbf{1.76} 	& 0.06 	& 0.21 	& \textbf{0.27} 	& 0.85\\
    \bottomrule
% \multicolumn{6}{l}{\footnotesize{We use $\rho \in[0,1]$ to denote the pruning rate.}}
  \end{tabular}
 \vspace{-12pt}
\end{table}

\subsection{Performance Evalutions}
\begin{figure*}\centering
  \includegraphics[width=0.872\textwidth]{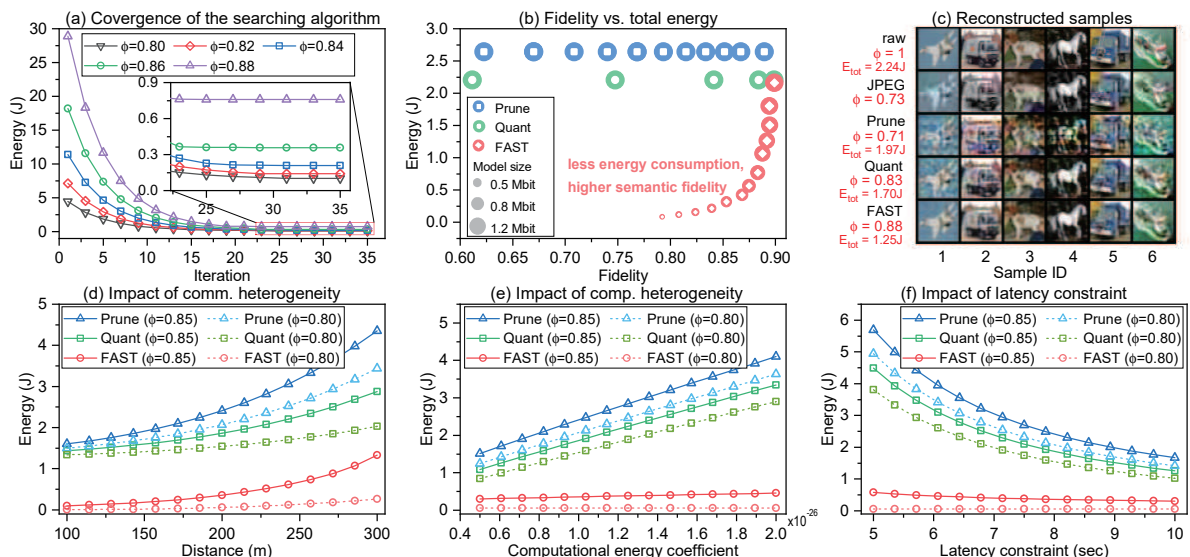}
  \vspace{-10pt}
  \caption{The main advantages of FAST. ((a): convergence of the searching algorithm; (b-c): performance of different methods; (d-f): impact of key settings.) 
}\label{fig:exp}
  \vspace{-15pt}
\end{figure*}

We first show that the hierarchical bisection search algorithm can converge to the optimal solution of Problem (P2). Figure~\ref{fig:exp}(a) presents the evolution of the total energy cost with respect to the number of iterations. We observe that the algorithm can achieve the optimum after about 30 iterations.

We next compare the proposed FAST with the following three baseline methods under $T^{\max}=8$ seconds.
{\bf  (1) JPEG}: we reduce the size of the raw data with a radical compression ratio of about 4.5.
{\bf (2) Prune}: we employ filter-wise feature pruning for the semantic data, and $\rho \in[0,1]$ is the pruning rate.
{\bf (3) Quant}: we quantize the semantic data with fewer bits (from 1 to 8 bits) before the transmission. 

Table \ref{tab:compare-exp} provides the comparison results of FAST against the baseline methods in terms of the size of semantic data, computation cost, and energy consumption under two types of fidelity constraints. 
% The advantage of FAST is that it can jointly save both communication and computing costs.
Compared with Prune and Quant, FAST can respectively reduce 15 times and 6.9 times the total energy consumption for realizing semantic communication with the fidelity of 0.80 and 0.85.  Particularly, the proposed FAST can reduce 13.7 times the data size under the low fidelity scenario.
Meanwhile, one of the important advantages of FAST is that the proposed flexible semantic model enables on-demand computation to mitigate the computation cost.

Figure~\ref{fig:exp}(b) shows the total energy consumption of different methods with different fidelity constraints. With a given fidelity constraint, the proposed FAST consistently outperforms the baseline methods to mitigate the total energy consumption. Specifically, FAST can switch to the small sub-model to significantly reduce the computation cost in the low-fidelity scenario. Figure~\ref{fig:exp}(c) provides the image plots of reconstructed samples with different fidelity constraints. Specifically, the proposed FAST can achieve the best semantic fidelity with the least total energy cost, which strikes the balance between transmission quality and resource utilization.

Figures \ref{fig:exp}(d)-(f) show how the energy efficiency is affected by the key system configurations, including the transmission distance (i.e., channel state), computational energy coefficient, and latency constraint. It can be observed that the proposed FAST consistently outperforms the baseline methods under a variety of system settings. The experiments show that FAST is more resilient than other baselines to achieve green transmission under heterogeneous scenarios.

% \subsection{Impact of Key Mechanisms and Hyper-parameters}

\section{Conclusion}
In this paper, we proposed FAST, a fidelity-adjustable semantic transmission framework for green communication. 
We presented the dynamic model training scheme to enable wireless devices to adopt different sub-model on demand.
To improve the energy efficiency of FAST, we focused on minimizing the total energy consumption under personalized latency and fidelity constraints.
By leveraging the theoretical analysis, we transformed the optimization problem into a tractable form and designed an algorithm to efficiently search for the optimal transmission strategy.
Experimental results demonstrate the advantage of FAST in improving energy efficiency and semantic quality against the baseline methods.

\section*{Acknowledgment}
Rong Yu and Yuan Wu are the corresponding authors. This work was supported in part by National Natural Science Foundation of China under Grants 61971148, U22A2054 and 62102099, in part by National Key R\&D Program of China under Grants 2020YFB1807802 and 2020YFB1807800, in part by Science and Technology Development Fund of Macau SAR under Grant 0162/2019/A3, in part by FDCT-MOST Joint Project under Grant 0066/2019/AMJ, and in part by Research Grant of University of Macau under Grant MYRG2020-00107-IOTSC. This work was also supported in part by the National Research Foundation (NRF), Singapore and Infocomm Media Development Authority under the Future Communications Research Development Programme (FCP), and DSO National Laboratories under the AI Singapore Programme (AISG Award No: AISG2-RP-2020-019).

\bibliographystyle{IEEEtran}
\bibliography{reference.bib}

\end{document}